
{

\font\titlefont=cmr10 scaled\magstep3

\nopagenumbers

\topskip200pt

\centerline{ \titlefont Symmetry Breaking in the Static
Coordinate System }
\centerline{ \titlefont of de Sitter Spacetime }

\vskip50pt

\centerline{
Andr\'as Kaiser\footnote{$^1$}
{andras@genesis2.physics.yale.edu}
and
Alan Chodos\footnote{$^2$}
{chodos@yalph2.physics.yale.edu}
}

\centerline{ Center for Theoretical Physics,
Sloane Physics Laboratory }
\centerline{ Yale University, New Haven, Connecticut
06520-8120 }

\vfill

\eject

}

{ \bf Abstract: }

\smallskip

We study symmetry breaking in the static
coordinate-system of de Sitter space. This is done
with the help of the functional-Schr\"odinger approach
used in previous calculations by Ratra [1]. We consider
a massless, minimally coupled scalar field as the
parameter of a continuous symmetry (the angular
component of an O(2) symmetry). Then we study the
correlation function of the massless scalar field, to
derive the correlation function of the original field,
which finally shows the restoration of the continuous
symmetry.

\bigskip

{ \bf 1.Introduction }

\smallskip

Studying quantum field theory in de Sitter space has a
long history [2], and the interest in the inflationary
model of the universe brought some attention to the
study of symmetry breaking in this spacetime [3].
As is known from previous results [1,4], the question
of whether an internal symmetry is spontaneously broken
in a field theory quantized on a de Sitter background
may depend on the coordinate system that is chosen. In
this paper, we choose to do the computation for a
minimally coupled scalar field in the so-called static
coordinate system, which is distinguished by the fact
that the interpretation of the results is particularly
transparent in this case.

The coordinate-system we wish to use has the special
property that its spatial sections are geodesic
hypersurfaces with identical metric, and the time
coordinate for each slice is given by the proper time
of the observer whose spatial coordinate is the origin.
In this sense this system is the one that describes
the spacetime as a geodesic observer perceives it. It
inherently excludes anything beyond the event horizon
of the observer, and manifests time-translational
invariance, because time-translation is generated by an
element of the de Sitter group (which is not the case
for the systems used previously), namely the one that
translates the observer along its world-line
($\Omega_{04}$, rotation in the 0-4 plane according to
the coordinate definitions
of section $2$). Only if the field is quantized in this
system, will the field modes have $e^{i{\omega}t}$
time-dependence, with the $t$ coordinate time being the
proper time of the observer at the origin. The
important consequence of this property is that this
quantization gives us the Fock-space as perceived by
the observer (i.e. the vacuum state of this
quantization will be the state in which the observer
doesn't detect particles, etc.). So if we want to
define symmetry breaking according to the field
measurements of our geodesic observer in its vacuum
state, then we must calculate the expectation value of
the field on the vacuum state of this quantization.

\bigskip

{ \bf 2.The static system }

\smallskip

De Sitter spacetime can be embedded in a 5D
Minkowski-space (we use signature  $<+,-,-,-,->$),
where it is a 4D hypersurface determined by

$$
x^{_02} - x^{_12} - x^{_22} - x^{_32} - x^{_42}
=
- \alpha^2
\eqno(2.1)
$$

where $\alpha$ is the de Sitter radius, which also
determines the spacetime curvature,
$R = 12 \alpha^{-2}$ [5].

\smallskip

For the static system we will use a time and polar
spatial coordinates, \ $t,r,\vartheta,\varphi$. \
Choosing the worldline of our geodesic observer to be

$$
x = \big( \ \alpha \sinh( t / \alpha ) \ ,
\, 0 \ , \, 0 \ , \, 0 \ ,
\, \alpha \cosh( t / \alpha ) \ \big)
\eqno(2.2)
$$

in terms of the coordinates of the embedding space
determines the geodesic 3-surfaces orthogonal to the
worldline at every $t$, which are given by the \
$x^{_0} = \tanh( t / \alpha ) \, x^{_4}$ \ condition
within the de Sitter manifold. These spatial
hypersurfaces meet on the 2-surface

$$
x^{_0} = 0 , \quad
x^{_12} + x^{_22} + x^{_32} = \alpha^2 , \quad
x^{_4} = 0
\eqno(2.3)
$$

This 2-surface thus forms a coordinate singularity in
the static system, which must be exluded from the range
of the coordinates. The physical reason for the
existence of this singularity is that null-geodesics
passing through it form the event horizon of our
observer. We will use

$$
r^2 = x^{_12} + x^{_22} + x^{_32}
\eqno(2.4)
$$

radial coordinate on the geodesic 3-surface of a
time-instant, even though this is not the proper
distance from the origin on the spatial sections.
The proper distance from the origin on these
hypersurfaces is given by \ \
$s = \alpha \arcsin( r / \alpha )$ . \ We also
introduce polar coordinates on the \ $r=const. \ \
t=const.$ \ 2-spheres in the usual way.

The embedding relations of the static coordinate system
are given by:

$$
\eqalign{
x^{_0} &= \alpha ( 1 - r^2 / \alpha^2 )^{ 1 / 2 }
\sinh( t / \alpha ) \cr
x^{_1} &= r \, \sin \vartheta \, \sin \varphi \cr
x^{_2} &= r \, \sin \vartheta \, \cos \varphi \cr
x^{_3} &= r \, \cos \vartheta \cr
x^{_4} &= \alpha ( 1 - r^2 / \alpha^2 )^{ 1 / 2 }
\cosh( t / \alpha ) \cr
}
\eqno(2.5)
$$

where the coordinate ranges are:

$$
t : \ ( -\infty , \infty ) \ , \qquad
r : \ [ 0 , \alpha ) \ , \qquad
\vartheta : \ [ 0 , \pi ] \ , \qquad
\varphi : \ [ 0 , 2 \pi ]
\eqno(2.6)
$$

As we can see from these relations, the static system
covers only a strip on the de Sitter manifold, for
which \ \ $x^{_12} + x^{_22} + x^{_32} < \alpha^2$ \ .
\ This is the set of points on the manifold, which are
in causal relation with the observer in both
directions.

The metric in this coordinate system is given by:

$$
ds^2
=
( 1 - r^2 / \alpha^2 ) dt^2 -
{1 \over 1 - r^2 / \alpha^2 } dr^2 -
r^2 \{ d\vartheta^2 +
\sin^2 \vartheta \, d\varphi^2 \}
\eqno(2.7)
$$

from which we can see that the observer at the origin
perceives an effect of time dilatation, i.e. time seems
to pass slower as one looks further away, an effect due
to the spacetime curvature (similarly to decreasing
radial coordinate near a black hole).

\smallskip

We write the action for a minimally coupled scalar
field $\phi$, with mass m:

$$
S
=
\int dt \, dr \, d\vartheta \, d\varphi \
r^2 \sin \vartheta \,
\bigg\{
{ { 1 \over 2 } { | \dot \phi | }^2
\over  1 - r^2 / \alpha^2 } -
( 1 - r^2 / \alpha^2 ) { 1 \over 2 }
\bigg|
{ \partial \phi  \over \partial r }
\bigg|^2 +
{ 1 \over 2 } \,
{ \phi^* { \cal L }_{ \vartheta \varphi }
( \phi ) \over r^2 } -
{ m^2 \over 2 } | \phi |^2
\bigg\}
\eqno(2.8)
$$

where ${ \cal L }_{ \vartheta \varphi }$ is the angular
part of the Laplacian at unit radius:

$$
{ \cal L }_{ \vartheta \varphi } =
{ \partial^2 \over \partial \vartheta^2 } +
\cot \vartheta
{ \partial \over \partial \vartheta } +
{ 1 \over \sin^2 \vartheta }
{ \partial^2 \over \partial \varphi^2 }
\eqno(2.9)
$$

This action leads to the wave equation:

$$
0
=
{ r^2 \over 1 - r^2 / \alpha^2 } \,
{ \partial ^2 \phi \over \partial t^2 } -
{ \partial \over \partial r }
\bigg( r^2 ( 1 - r^2 / \alpha^2 )
{ \partial \phi  \over \partial r } \bigg) -
{ \cal L }_{ \vartheta \varphi } ( \phi ) +
r^2 m^2 \phi
\eqno(2.10)
$$

Notice that by redefining of the radial coordinate

$$
\tilde{r} =
{ 1 \over 2 } \, \alpha \,
\ln
\bigg( { \alpha + r \over \alpha - r } \bigg)
\qquad \quad
r = \alpha \tanh( \tilde{r} / \alpha )
$$
$$
d \tilde{r} = { dr \over 1 - r^2 / \alpha^2 }
\qquad \quad
dr = { d \tilde{r} \over
\cosh^2( \tilde{r} / \alpha ) }
\eqno(2.11)
$$
$$
\tilde{r} : \ [ 0 , \infty )
\qquad \quad
r : \ [ 0 , \alpha )
$$

we can bring the action to the following form:

$$
S
= \int
dt \, d\tilde{r} \, d\vartheta \, d\varphi \
\big(
\alpha \tanh( \tilde{r} / \alpha )
\big)^2 \,
\sin \vartheta \,
\bigg\{
{ 1 \over 2 } { | \dot \phi | }^2  -
{ 1 \over 2 }
\bigg|
{ \partial \phi  \over \partial \tilde{r} }
\bigg|^2 +
{ 1 \over 2 \cosh^2( \tilde{r} / \alpha ) } \,
{ \phi^*
{ \cal L }_{ \vartheta \varphi } ( \phi )
\over
\big(
\alpha \tanh( \tilde{r} / \alpha )
\big)^2 }  -
{ m^2 \over 2 \cosh^2( \tilde{r} / \alpha ) }
| \phi |^2
\bigg\}
\eqno(2.12)
$$

and the wave equation:

$$
0
=
\Big(
\alpha \tanh( \tilde{r} / \alpha )
\Big)^2 \
{ \partial ^2 \phi \over \partial t^2 } -
{ \partial \over \partial \tilde{r} }
\bigg(
\Big(
\alpha \tanh( \tilde{r} / \alpha )
\Big)^2
{ \partial \phi  \over \partial \tilde{r} }
\bigg)
-
{ { \cal L }_{ \vartheta \varphi } ( \phi )
\over \cosh^2( \tilde{r} / \alpha ) } +
{ m^2 \phi \over \cosh^2( \tilde{r} / \alpha ) }
\eqno(2.13)
$$

For $\tilde{r} / \alpha \gg 1$ the action becomes

$$
S
=
\int
dt \, d\tilde{r} \, d\vartheta \, d\varphi \
\alpha^2 \sin \vartheta \,
\bigg\{
{ 1 \over 2 } { | \dot \phi | }^2  -
{ 1 \over 2 }
\bigg|
{ \partial \phi  \over \partial \tilde{r} }
\bigg|^2
\bigg\}
\eqno(2.14)
$$

and the wave equation

$$
0
=
\alpha^2 \
{ \partial ^2 \phi \over \partial t^2 } -
{ \partial \over \partial \tilde{r} }
\bigg(
\alpha^2
{ \partial \phi  \over \partial \tilde{r} }
\bigg)
=
\alpha^2 \
\bigg(
{ \partial ^2 \phi \over \partial t^2 } -
{ \partial ^2 \phi  \over \partial \tilde{r}^2 }
\bigg)
\eqno(2.15)
$$

which yields plane waves with respect to the
$\tilde{r}$ conformal radial coordinate in this limit.
Looking at the relation between $r$ and $\tilde{r}$
again, we can check that the conformal coordinate of a
point is given by the coordinate time that is needed
for a light ray to travel between the origin and the
point in question. From this we can also see that the
plane wave gets infinitely compressed
($ d\tilde{r} / dr \rightarrow \infty $) as it
approaches the event horizon, which is due to the time
dilatation discussed previously, i.e. it takes infinite
coordinate time for a light ray to reach the event
horizon.

\bigskip

Now return to the original (non-conformal) radial
parameter $r$, and look for solutions in the form:

$$
\Phi
=
e^{-i \omega t} \, \hat\phi_{ \omega l m }
( r , \vartheta , \varphi )
=
e^{-i \omega t} \, f_{\omega l} ( r ) \,
Y^m_l ( \vartheta , \varphi )
\eqno(2.16)
$$

The solutions are [6]:

$$
\eqalign{
&
\hat\phi_{ \omega l m }( r,\vartheta,\varphi )
=
\cr
&
\cr
&
=
{
\Gamma\Big( { l \over 2 } + { 3 \over 4 } +
{ i \omega \alpha \over 2 } + { 1 \over 2 }
\sqrt{ { 9 \over 4 } - m^2 \alpha^2 } \ \Big) \,
\Gamma\Big( { l \over 2 } + { 3 \over 4 } +
{ i \omega \alpha \over 2 } - { 1 \over 2 }
\sqrt{ { 9 \over 4 } - m^2 \alpha^2 } \ \Big) \,
\over
\sqrt{2 \pi} \,
\Gamma( l + { 3 \over 2 } ) \,
\Gamma( i \omega \alpha ) \,
} \
\bigg( { r^2 \over \alpha^2 } \bigg)^{ l/2 }
( 1 - r^2 / \alpha^2 )^
{ i \omega \alpha \over 2 } \,
\cr
&
F \bigg( \,
{ l \over 2 } + { 3 \over 4 } +
{ i \omega \alpha \over 2 } + { 1 \over 2 }
\sqrt{ { 9 \over 4 } - m^2 \alpha^2 }
\ \ ,\ \
{ l \over 2 } + { 3 \over 4 } +
{ i \omega \alpha \over 2 } - { 1 \over 2 }
\sqrt{ { 9 \over 4 } - m^2 \alpha^2 }
\ \ ; \ \
l + { 3 \over 2 }
\ \ ; \ \
{ r^2 \over \alpha^2 } \,
\bigg)
Y_l^m ( \vartheta , \varphi )
\cr
}
$$
$$
\eqno(2.17)
$$

where $F$ is the hypergeometric function. The solutions
are orthonormalized with the following measure:

$$
\int dr \, d\vartheta \, d\varphi \
{ r^2 \sin( \vartheta ) \over 1 - r^2 / \alpha^2 } \
\hat\phi^*_{ \omega' l' m' }
( r,\vartheta,\varphi ) \
\hat\phi_{ \omega l m }
( r,\vartheta,\varphi )
=
\delta( \omega' , \omega ) \,
\delta_{ l' , l } \,
\delta_{ m' , m }
\eqno(2.18)
$$

We can expand the field in terms of these modes:

$$
\phi( r , \vartheta , \varphi )
=
\sum_{ l , m } \int d\omega \
\xi_{ \omega l m } \,
\hat\phi_{ \omega l m }
( r,\vartheta,\varphi )
\eqno(2.19)
$$

then express the action:

$$
S=\sum_{ l , m } \int dt \, d\omega \,
\bigg\{
{ 1 \over 2 } | \dot \xi_{ \omega l m } |^2 -
{ \omega^2 \over 2 } | \xi_{ \omega l m } |^2
\bigg\}
\eqno(2.20)
$$

We have the action of a simple harmonic oscillator for
each mode, which has the ground-state wavefunction:

$$
\psi( \xi_{ \omega l m } \, ,t ) =
e^{ - { i \omega t \over 2 } }
\bigg( { \omega \over \pi } \bigg)
^{ 1 \over 4 }
e^{ - { \omega \over 2 } \xi^2 }
\eqno(2.21)
$$

\vbox{

from which immediately follows:

$$
< \xi^2_{ \omega l m } > = { 1 \over 2\omega }
\eqno(2.22)
$$

}

\bigskip

{ \bf 3.The correlation functions }

\smallskip

Using this, and the field expansion, we can easily
express the equal-time 2-point function:

$$
< \phi( r',\vartheta',\varphi' ) \, ,\,
\phi( r,\vartheta,\varphi ) >
=
\sum_{ l,m } \int { d\omega \over 2 \omega } \,
\hat\phi^*_{ \omega l m }
( r',\vartheta',\varphi' ) \,
\hat\phi_{ \omega l m }
( r,\vartheta,\varphi )
\eqno(3.1)
$$

If we set \ $r'=0$ , the \ $l \ne 0$ \ modes can be
neglected, since they are zero at the origin. We don't
have to care about $\vartheta$ or $\varphi$ either,
because of the spherical symmetry of the \ $l=0$ \
modes.

$$
< \phi( 0 ) \, ,\, \phi( r ) > =
\int { d\omega \over 2 \omega } \,
\hat\phi^*_{ \omega 0 0 }( 0 ) \,
\hat\phi_{ \omega 0 0 }( r )
=
$$
$$
=
\int { d\omega \over 2\omega } \,
{
\Gamma\Big(
{ 3 \over 4 } - { i \omega \alpha \over 2 } +
{ 1 \over 2 }
\sqrt{ { 9 \over 4 } - m^2 \alpha^2 } \
\Big) \,
\Gamma\Big(
{ 3 \over 4 } - { i \omega \alpha \over 2 } -
{ 1 \over 2 }
\sqrt{ { 9 \over 4 } - m^2 \alpha^2 } \
\Big) \,
\over
\sqrt{ 2 \pi } \,
\sqrt{ 4 \pi } \,
\Gamma( { 3 \over 2 } ) \,
\Gamma( -i \omega \alpha  ) \,
}
\eqno(3.2)
$$
$$
{
\Gamma\Big(
{ 3 \over 4 } + { i \omega \alpha \over 2 } +
{ 1 \over 2 }
\sqrt{ { 9 \over 4 } - m^2 \alpha^2 } \
\Big) \,
\Gamma\Big(
{ 3 \over 4 } + { i \omega \alpha \over 2 } -
{ 1 \over 2 }
\sqrt{ { 9 \over 4 } - m^2 \alpha^2 } \
\Big) \,
\over
\sqrt{ 2 \pi } \,
\sqrt{ 4 \pi } \,
\Gamma ( { 3 \over 2 } ) \,
\Gamma ( i \omega \alpha ) \,
} \,
( 1 - r^2 / \alpha^2  )^
{ i \omega \alpha \over 2 } \,
$$
$$
F
\bigg( \,
{ 3 \over 4 } + { i \omega \alpha \over 2 } +
{ 1 \over 2 }
\sqrt{ { 9 \over 4 } - m^2 \alpha^2 }
\ \ ,\ \
{ 3 \over 4 } + { i \omega \alpha \over 2 } -
{ 1 \over 2 }
\sqrt{ { 9 \over 4 } - m^2 \alpha^2 }
\ \ ; \ \
{ 3 \over 2 }
\ \ ; \ \
r^2 / \alpha^2  \,
\bigg)
$$

Examine the \ $r=0$ \ case first. The hypergeometric
function gives $1$ at zero argument, and since \ \
${ \Gamma( iy ) \Gamma( -iy ) = { \pi \over y \,
\sinh ( \pi y ) } }$ , \ we have an expression with a
powerlike ultraviolet divergence. In the massless case
we also have an infrared divergence, due to the square
root cancelling the $3/4$ in the argument of the gamma
function.

Now examine \ $r \rightarrow \alpha$ \ ( $r = \alpha$ \
is the event horizon, as discussed in section $2$). Use
the following transformation of the hypergeometric
function [10]:

$$
F ( a \ , \ b \ ; \ c \ ; \ z )
=
$$
$$
{ \Gamma( c ) \, \Gamma( c - a - b )
\over \Gamma( c - a ) \, \Gamma( c - b ) } \
F ( a \ , \ b \ ;
\ a + b - c + 1 \ ; \ 1 - z )
\ \ +
\eqno(3.3)
$$
$$
+ \ \
( 1 - z )^{ c - a - b } \
{ \Gamma( c ) \, \Gamma( a + b - c )
\over \Gamma( a ) \, \Gamma( b ) } \
F ( c - a \ , \ c - b \ ;
\ c - a - b + 1 \ ; \ 1 - z )
$$

which gives:

$$
< \phi( 0 ) \, ,\, \phi( r ) >
=
$$
$$
=
\int { d\omega \over 16 \omega \pi^2 } \,
\ \
\Bigg\{ \ \ \ \
{
\Gamma\Big(
{ 3 \over 4 } + { i \omega \alpha \over 2 } +
{ 1 \over 2 } \sqrt{ { 9 \over 4 } -
m^2 \alpha^2 } \
\Big) \,
\Gamma\Big(
{ 3 \over 4 } + { i \omega \alpha \over 2 } -
{ 1 \over 2 } \sqrt{ { 9 \over 4 } -
m^2 \alpha^2 } \
\Big) \,
\over
\Gamma ( { 3 \over 2 } ) \,
\Gamma ( i \omega \alpha ) \,
} \,
( 1 - r^2 / \alpha^2 )^
{ i \omega \alpha \over 2 } \,
$$
$$
F
\bigg( \,
{ 3 \over 4 } + { i \omega \alpha \over 2 } +
{ 1 \over 2 }
\sqrt{ { 9 \over 4 } - m^2 \alpha^2 }
\ \ ,\ \
{ 3 \over 4 } + { i \omega \alpha \over 2 } -
{ 1 \over 2 }
\sqrt{ { 9 \over 4 } - m^2 \alpha^2 }
\ \ ; \ \
i \omega \alpha + 1
\ \ ; \ \
1 - r^2 / \alpha^2 \,
\bigg)
\ \ +
\eqno(3.4)
$$
$$
+ \ \
{
\Gamma\Big(
{ 3 \over 4 } - { i \omega \alpha \over 2 } +
{ 1 \over 2 } \sqrt{ { 9 \over 4 } -
m^2 \alpha^2 } \
\Big) \,
\Gamma\Big(
{ 3 \over 4 } - { i \omega \alpha \over 2 } -
{ 1 \over 2 } \sqrt{ { 9 \over 4 } -
m^2 \alpha^2 } \
\Big) \,
\over
\Gamma ( { 3 \over 2 } ) \,
\Gamma ( -i \omega \alpha ) \,
} \,
( 1 - r^2 / \alpha^2 )^
{ -i \omega \alpha \over 2 } \,
$$
$$
F
\bigg( \,
{ 3 \over 4 } + { -i \omega \alpha \over 2 } +
{ 1 \over 2 }
\sqrt{ { 9 \over 4 } - m^2 \alpha^2 }
\ \ ,\ \
{ 3 \over 4 } + { -i \omega \alpha \over 2 } -
{ 1 \over 2 }
\sqrt{ { 9 \over 4 } - m^2 \alpha^2 }
\ \ ; \ \
-i \omega \alpha + 1
\ \ ; \ \
1 - r^2 / \alpha^2 \,
\bigg) \ \
\Bigg\}
$$

As \ $r \rightarrow \alpha$ , the arguments of the
hypergeometric functions approach zero, therefore their
values approach~$1$, so we set them to $1$ for the
purposes of this discusssion. In the next section we
will need to know the behaviour of some of these
expressions near the \ $r = \alpha$ \ limit, having the
freedom to choose the radial coordinate arbitrarily
close to $\alpha$. In the following analysis we will
have to make some approximations to be able to
investigate certain properties of this expression.
These approximations will cause some finite error in
the result, but it will not be crucial for two reasons.
The first reason is that we will study certain
divergences in the obtained expressions, for which the
finite error is irrelevant. The second reason is that
the error can be made arbitrarily small by choosing the
radial coordinate sufficiently close to $\alpha$, when
we evaluate the expressions.

On the other hand, the \ \ ${ ( 1 - r^2 / \alpha^2 )^
{ \pm { i \omega \alpha \over 2 } } }$ \ \ term can be
expressed as \ \ ${ e^{ \pm i \ln( 1 - r^2 / \alpha^2 )
{ \omega \alpha \over 2 } } }$ , \ which results in an
infinitely fast oscillation of this term with respect
to $\omega$, as \ $r \rightarrow \alpha$ .

Focus on the \ $m \rightarrow 0$ \ case, taking the \
$m \rightarrow 0$ \ limit before the \ $r \rightarrow
\alpha$ \ limit, because this is the case that we will
need in the next section. As \ $m \rightarrow 0$ , we
can substitute \ \ $\Gamma\big( { 3 \over 2 } - { m^2
\alpha^2 \over 6 } \pm { i \omega \alpha \over 2 }
\big)$ \ \ for \ \ ${ \Gamma\Big( { 3 \over 4 } \pm
{ i \omega \alpha \over 2 } + { 1 \over 2 }
\sqrt{ { 9 \over 4 } - m^2 \alpha^2 } \ \Big) }$
, \ \ and \ \ \
$\Gamma\big( { m^2 \alpha^2 \over 6 }
\pm { i \omega \alpha \over 2 } \big)$ \ \ \
for \ \ \
${ \Gamma\Big( { 3 \over 4 } \pm { i \omega \alpha
\over 2 } - { 1 \over 2 } \sqrt{ { 9 \over 4 } - m^2
\alpha^2 } \ \Big) }$ . \ As we take the \
$m \rightarrow 0$ \ limit, the error resulting from
this substitution will also go to zero.

\smallskip

After these substitutions we have:

$$
\int { d\omega \over 16 \omega \pi^2 } \,
\
\Bigg\{ \ \
{
\Gamma\big( { 3 \over 2 } -
{ m^2 \alpha^2 \over 6 } +
{ i \omega \alpha \over 2 } \ \big) \,
\Gamma\big( { m^2 \alpha^2 \over 6 } +
{ i \omega \alpha \over 2 } \ \big) \,
\over
\Gamma ( { 3 \over 2 } ) \,
\Gamma ( i \omega \alpha ) \,
} \,
( 1 - r^2 / \alpha^2 )^
{ i \omega \alpha \over 2 }
\ \ +
$$
$$
+ \ \
{
\Gamma\big( { 3 \over 2 } -
{ m^2 \alpha^2 \over 6 } -
{ i \omega \alpha \over 2 } \ \big) \,
\Gamma\big( { m^2 \alpha^2 \over 6 } -
{ i \omega \alpha \over 2 } \ \big) \,
\over
\Gamma ( { 3 \over 2 } ) \,
\Gamma ( -i \omega \alpha ) \,
} \,
( 1 - r^2 / \alpha^2 )^
{ -i \omega \alpha \over 2 } \ \
\Bigg\}
\eqno(3.5)
$$

Now break up the $\omega$ integral into three parts:

$$
\eqalign{
0 \quad
&\hbox{---} \quad
m^2 \alpha / \, 3 \cr
m^2 \alpha / \, 3 \quad
&\hbox{---} \quad
{ 2 \pi / ( - \alpha \ln( 1 - r^2 / \alpha^2 ) ) } \cr
{ 2 \pi / ( - \alpha \ln( 1 - r^2 / \alpha^2 ) ) }
\quad
&\hbox{---} \quad
\infty \cr
}
\eqno(3.6)
$$

Taking the \ $m \rightarrow 0$ \ limit faster than the
\ $r \rightarrow \alpha$ \ limit assures this order and
the \ \ ${ m^2 \alpha / 3 \ll }$ ${ 2 \pi / ( - \alpha
\ln( 1 - r^2 / \alpha^2 ) }$ \ \ condition.

\medskip

Consider the first part. The above condition allows us
to substitute $1$ for \ \ ${ ( 1 - r^2 / \alpha^2 )^
{ \pm { i \omega \alpha \over 2 } } }$ . \ As the \ $m
\rightarrow 0$ \ limit is taken faster than the \ $r
\rightarrow \alpha$ \ limit, \ \ $\omega \ln( 1 - r^2 /
\alpha^2 )$ \ \ goes to zero, justifying the above
approximation. Also, all the errors that the following
approximations will cause go to zero with \ $m
\rightarrow 0$ . Substitute \ \ $\Gamma( { 3 \over 2 }
)$ \ \ for \ \ ${ \Gamma( { 3 \over 2 } - { m^2
\alpha^2 \over 6 } \pm { i \omega \alpha \over 2 } ) }$
, \ since \ \ $m \rightarrow 0$ , \ and \ \ ${ \omega
\alpha \over 2 } \le { m^2 \alpha^2 \over 6 }$ \ \ in
this interval, and the gamma function does not have a
pole at $3 / 2$.  Then substitute \ $1 / z$ \ for \
$\Gamma( z )$ , wherever \ $|z| \ll 1$ , which gives us

$$
\int\limits_0^{ m^2 \alpha / 3 }
{ d\omega \over 16 \omega \pi^2 } \
\bigg\{ \
{
i \omega \alpha
\over
{ m^2 \alpha^2 \over 6 } +
{ i \omega \alpha \over 2 }
}
\ + \
{
-i \omega \alpha
\over
{ m^2 \alpha^2 \over 6 } -
{ i \omega \alpha \over 2 }
}
\
\bigg\}
\eqno(3.7)
$$

which shows that this first of the three parts goes to
a (finite) constant as $m \rightarrow 0$.

\medskip

Next consider the third interval of integration, \quad
$\omega: \quad { 2 \pi / ( - \alpha \ln( 1 - r^2 /
\alpha^2 ) ) } \quad \hbox{---} \quad \infty$ . \quad
Here we can replace \ \ ${ \Gamma( { 3 \over 2 } - {
m^2 \alpha^2 \over 6 } \pm { i \omega \alpha \over 2 }
) }$ \ \ \ with \ \ \ ${ \Gamma( { 3 \over 2 } \pm { i
\omega \alpha \over 2 } ) }$ \ \ , and \ \ \ ${ \Gamma(
{ m^2 \alpha^2 \over 6 } \pm { i \omega \alpha \over 2
} ) }$ \ \ \ with \ \ \ $\Gamma( \pm { i \omega \alpha
\over 2 } )$ \ \ \ in the \ \ \ $m \rightarrow 0$ \ \
limit. Then use \ \ ${ \Gamma( \pm i \omega \alpha ) =
( 2 \pi )^{ - { 1 \over 2 } } 2^{ \pm i \omega \alpha -
{ 1 \over 2 } } \Gamma( { 1 \over 2 } \pm { i \omega
\alpha \over 2 } ) \Gamma( \pm { i \omega \alpha \over
2 } ) }$ \ \

\vbox{

This way we get

$$
\int\limits_
{ 2 \pi / ( - \ln( 1 - r^2 / \alpha^2 ) ) }^
{ \infty }
{ d\omega \over 16 \omega \pi^2 } \,
\
\Bigg\{ \ \
{
\Gamma\big(
{ 3 \over 2 } + { i \omega \alpha \over 2 }
\big) \,
\Gamma\big(
{ i \omega \alpha \over 2 }
\big) \,
\over
\Gamma ( { 3 \over 2 } ) \,
( 2 \pi )^{ - { 1 \over 2 } } \,
2^{ i \omega \alpha - { 1 \over 2 } } \,
\Gamma\big(
{ 1 \over 2 } + { i \omega \alpha \over 2 }
\big) \,
\Gamma\big(
{ i \omega \alpha \over 2 }
\big) \,
} \,
( 1 - r^2 / \alpha^2 )^
{ i \omega \alpha \over 2 }
\ \ +
$$
$$
+ \ \
{
\Gamma\big(
{ 3 \over 2 } - { i \omega \alpha \over 2 }
\big) \,
\Gamma\big(
{ -i \omega \alpha \over 2 }
\big) \,
\over
\Gamma ( { 3 \over 2 } ) \,
( 2 \pi )^{ - { 1 \over 2 } } \,
2^{ -i \omega \alpha - { 1 \over 2 } } \,
\Gamma\big(
{ 1 \over 2 } - { i \omega \alpha \over 2 }
\big) \,
\Gamma\big(
{ -i \omega \alpha \over 2 }
\big) \,
} \,
( 1 - r^2 / \alpha^2 )^
{ -i \omega \alpha \over 2 } \ \
\Bigg\}
\eqno(3.8)
$$

}

then use \ \ $\Gamma( z + 1 ) = z \Gamma( z )$ , \
cancel the appropriate terms, and rearrange:

$$
\int\limits_
{ 2 \pi / ( - \alpha
\ln( 1 - r^2 / \alpha^2 ) ) }^{ \infty }
{ d\omega \over 16 \omega \pi^2 } \,
\
\Bigg\{ \ \
{
( 1 + i \omega \alpha ) \,
\sqrt{ \pi } \,
\over
\Gamma ( { 3 \over 2 } ) \,
} \,
\bigg( { 1 - r^2 / \alpha^2 \over 4 } \bigg)^
{ i \omega \alpha \over 2 }
\ \ + \ \
{
( 1 - i \omega \alpha ) \,
\sqrt{ \pi } \,
\over
\Gamma ( { 3 \over 2 } ) \,
} \,
\bigg( { 1 - r^2 / \alpha^2 \over 4 } \bigg)^
{ -i \omega \alpha \over 2 }
\ \ \Bigg\}
\eqno(3.9)
$$

break it into two integrals:

$$
\int\limits_
{ 2 \pi / ( - \alpha \ln( 1 - r^2 / \alpha^2 ) ) }^
{ \infty }
{ d\omega \over 16 \omega \pi^2 } \,
\
\Bigg\{ \ \
{
\sqrt{ \pi } \,
\over
\Gamma ( { 3 \over 2 } ) \,
} \,
\bigg( { 1 - r^2 / \alpha^2 \over 4 } \bigg)^
{ i \omega \alpha \over 2 }
\ \ + \ \
{
\sqrt{ \pi } \,
\over
\Gamma ( { 3 \over 2 } ) \,
} \,
\bigg( { 1 - r^2 / \alpha^2 \over 4 } \bigg)^
{ -i \omega \alpha \over 2 }
\ \ \Bigg\}
\ \ +
$$

$$
+ \ \
\int\limits_
{ 2 \pi / ( - \alpha \ln( 1 - r^2 / \alpha^2 ) ) }^
{ \infty }
{ d\omega \over 16 \pi^2 } \,
\
\Bigg\{ \ \
{
i \alpha \,
\sqrt{ \pi } \,
\over
\Gamma ( { 3 \over 2 } ) \,
} \,
\bigg( { 1 - r^2 / \alpha^2 \over 4 } \bigg)^
{ i \omega \alpha \over 2 }
\ \ + \ \
{
- i \alpha \,
\sqrt{ \pi } \,
\over
\Gamma ( { 3 \over 2 } ) \,
} \,
\bigg( { 1 - r^2 / \alpha^2 \over 4 } \bigg)^
{ -i \omega \alpha \over 2 }
\ \ \Bigg\}
\eqno(3.10)
$$

where the first integral is finite because of the
following. As a consequence of the oscillation, the
integral will consist of contributions with
periodically alternating sign, and decreasing
magnitude. Therefore the first contribution gives an
upper bound to the integral. We get an upper bound to
this first contribution, if we multiply the length of
the interval (half period, \ \ ${ 2 \pi / [ \alpha ( -
\ln{ 1 - r^2 / \alpha^2 \over 4 } ) ] }$ \ \ ), with
the supremum of the function in the interval, which
gives

$$
{ 2 \pi \over \alpha
( - \ln{ 1 - r^2 / \alpha^2 \over 4 } ) }
\
{ \sqrt{ \pi } \ / \ \Gamma ( { 3 \over 2 } )
\over 16 { 2 \pi \over - \alpha
\ln( 1 - r^2 / \alpha^2 ) } \pi^2 }
=
{ \Gamma ( { 3 \over 2 } )
\over
16 \pi \sqrt{ \pi } } \
{ \ln( 1 - r^2 / \alpha^2 )
\over
\ln( 1 - r^2 / \alpha^2 ) - \ln( 4 ) }
\eqno(3.11)
$$

which is finite ( \ \ $0 < r < \alpha \, , \ r
\rightarrow \alpha$ \ \ )

\smallskip

The second integral is the integral of a periodic
oscillating function, which we regulate with an \ \
$e^{ - \epsilon \omega }$ \ \ type ultraviolet cutoff,
then take \ \ $\epsilon \rightarrow 0$ \ \ limit,
obtaining

$$
\int\limits_
{ 2 \pi / ( - \alpha \ln( 1 - r^2 / \alpha^2 ) ) }^
{ \infty }
{ \alpha \over 16 \pi \sqrt{ \pi } \
\Gamma ( { 3 \over 2 } ) }
\
\Bigg\{ \ \
i \
e^{ -i \, \big( \,
\ln( { 4 \over 1 - r^2 / \alpha^2 } ) \
{ \omega \alpha \over 2 } \ - i \epsilon \omega \, \big)  }
\ \ - \ \
i \
e^{ i \, \big( \,
\ln( { 4 \over 1 - r^2 / \alpha^2 } ) \
{ \omega \alpha \over 2 } \ + i \epsilon \omega \, \big) }
\ \ \Bigg\}
=
$$

$$
=
{ \alpha \over 16 \pi \sqrt{ \pi } \
\Gamma ( { 3 \over 2 } ) }
\ \
\Bigg[ \
{ i \over -i \big(
\ln( { 4 \over 1 - r^2 / \alpha^2 } ) \
{ \alpha \over 2 } - i \epsilon \big) } \
e^{ -i \, \big( \,
\ln( { 4 \over 1 - r^2 / \alpha^2 } ) \
{ \omega \alpha \over 2 } \ - i \epsilon \omega \, \big)  }
\ \ +
$$
$$
+ \ \
{ -i \over i \big(
\ln( { 4 \over 1 - r^2 / \alpha^2 } ) \
{ \alpha \over 2 } + i \epsilon \big) } \
e^{ i \, \big( \,
\ln( { 4 \over 1 - r^2 / \alpha^2 } ) \
{ \omega \alpha \over 2 } \ + i \epsilon \omega \, \big) }
\ \Bigg]
_{ 2 \pi / ( - \alpha \ln( 1 - r^2 / \alpha^2 ) ) }
^{ \infty }
=
\eqno(3.12)
$$

$$
=
{ \alpha \over 16 \pi \sqrt{ \pi } \
\Gamma ( { 3 \over 2 } ) }
\ \
\Bigg( \
{ 1 \over
\ln( { 4 \over 1 - r^2 / \alpha^2 } ) \
{ \alpha \over 2 } - i \epsilon } \
e^{ -i \, \big( \, \pi \, {
\ln( 1 - r^2 / \alpha^2 ) - \ln( 4 ) \
\over \ln( 1 - r^2 / \alpha^2 )
} \ - { i \epsilon 2 \pi
\over - \alpha \ln( 1 - r^2 / \alpha^2 ) } \, \big)  }
\ \ +
$$
$$
+ \ \
{ 1 \over
\ln( { 4 \over 1 - r^2 / \alpha^2 } ) \
{ \alpha \over 2 } + i \epsilon } \
e^{ i \, \big( \, \pi \, {
\ln( 1 - r^2 / \alpha^2 ) - \ln( 4 ) \
\over \ln( 1 - r^2 / \alpha^2 )
} \ + { i \epsilon 2 \pi
\over - \alpha \ln( 1 - r^2 / \alpha^2 ) } \, \big) }
\ \Bigg)
$$

which is, once again, finite, and in the $\epsilon
\rightarrow 0$ limit yields

$$
{ 1 \over 4 \pi \sqrt{ \pi } \
\Gamma ( { 3 \over 2 } ) \
\ln( { 4 \over 1 - r^2 / \alpha^2 } ) \
} \
\cos\Big( \, \pi \ {
\ln( 1 - r^2 / \alpha^2 ) - \ln( 4 ) \
\over \ln( 1 - r^2 / \alpha^2 )
} \, \Big)
\eqno(3.13)
$$

from which we can see that this interval also gives
a finite contribution to the total integral.

\medskip

Now consider the second interval of integration, \quad
$\omega: \quad m^2 \alpha / 3 \quad \hbox{---} \quad
{ 2 \pi / ( - \alpha \ln( 1 - r^2 / \alpha^2 ) ) }$ .
\quad Here we can neglect the oscillation of the \ \
${ ( 1 - r^2 / \alpha^2 )^{ \pm { i \omega \alpha \over
2 } } }$ \ \ term without changing the magnitude of the
result, because the upper limit of integration was
determined such that the phase change of this term
always stays less than $\pi$. Also we can substitute \
\ $\Gamma( { 3 \over 2 } )$ \ \ \ for \ \ \ ${ \Gamma(
{ 3 \over 2 } - { 1 \over 6 } m^2 \alpha^2 \pm { i
\omega \alpha \over 2 } ) }$ , \ \ because \ \ \ ${ m^2
\alpha / 3 \ll 2 \pi / ( - \alpha \ln( 1 - r^2 /
\alpha^2 ) ) }$ , \ \ and from \ \ \ $r \approx \alpha$
\ \ \ follows \ \ \ ${ 2 \pi / - \ln( 1 - r^2 /
\alpha^2 ) \ll 1 }$ , \ \ so we have both \ \ \ ${ m^2
\alpha^2 \over 6 } \ll 1$ \ \ \ and \ \ \ ${ \omega
\alpha \over 2 } \ll 1$ . \ From this also follows that
we can use the \ $1 / z$ \ substitution for \ $\Gamma(
z )$ \ \ for the \ \ $\Gamma( { m^2 \alpha^2 \over 6 }
\pm { i \omega \alpha \over 2 } )$ \ \ and \ \ $\Gamma(
\pm i \omega \alpha )$ \ \ terms.

As we see, we finally arrive at the same expression
as in the first part, but here we have different limits
for the integration.

$$
\int\limits_{ m^2 \alpha / 3 }^
{ 2 \pi / ( - \alpha \ln( 1 - r^2 / \alpha^2 ) ) }
{ d\omega \over 16 \omega \pi^2 } \
\bigg\{ \
{
i \omega \alpha
\over
{ m^2 \alpha^2 \over 6 } +
{ i \omega \alpha \over 2 }
}
\ + \
{
-i \omega \alpha
\over
{ m^2 \alpha^2 \over 6 } -
{ i \omega \alpha \over 2 }
}
\
\bigg\}
$$
$$
=
\int\limits_{ m^2 \alpha / 3 }^
{ 2 \pi / ( - \alpha \ln( 1 - r^2 / \alpha^2 ) ) }
{ d\omega \over 16 \omega \pi^2 } \
{
( \omega \alpha )^2
\over
( { m^2 \alpha^2 \over 6 } )^2 +
( { \omega \alpha \over 2 } )^2
}
$$
$$
=
\int\limits_{ m^2 \alpha / 3 }
^{ 2 \pi / ( - \alpha \ln( 1 - r^2 / \alpha^2 ) ) }
{ d\omega \over 8 \pi^2 } \
{
\alpha \ { \omega \alpha \over 2 }
\over
( { m^2 \alpha^2 \over 6 } )^2 +
( { \omega \alpha \over 2 } )^2
}
\eqno(3.14)
$$
$$
=
{ 1 \over 8 \pi^2 }
\ln\Bigg(
\bigg( { \omega \alpha \over 2 } \bigg)^2 +
\bigg( { m^2 \alpha^2 \over 6 } \bigg)^2
\Bigg)
\Bigg|_{ \omega = m^2 \alpha / 3 }
^{ 2 \pi / ( - \alpha \ln( 1 - r^2 / \alpha^2 ) ) }
$$
$$
=
{ 1 \over 8 \pi^2 }
\ln\Bigg(
{
\big( { \pi \over -
\ln( 1 - r^2 / \alpha^2 ) } \big)^2 +
\big( { m^2 \alpha^2 \over 6 } \big)^2
\over
2 \, \big( { m^2 \alpha^2 \over 6 } \big)^2
}
\Bigg)
$$

which is a divergent expression in the \ $m
\rightarrow 0$ \ limit. We will need this expression in
the next section.

\bigskip

{ \bf 4.Symmetry breaking }

\smallskip

Now apply these results to the problem of symmetry
breaking. Consider a model  with complex scalar field
$\Phi$ and a symmetry-breaking $\Phi^4$ potential.
We will use the approximation of a fixed radial
component of the complex scalar field at the classical
minimum, \ $|\Phi| = \rho_0$ , which leaves us with a
massless  minimally coupled scalar field as the angular
component. We give a summary of this treatment. For a
fuller discussion see  [1].

In a broken-symmetry vacuum state the \ \ $< \Phi( 0 )
\, \Phi( r ) >$ \ \ correlation does not go to zero as
we let $r$ approach the event horizon, so a vanishing
correlation is evidence for symmetry restoration.
Express \ \ $< \Phi( 0 ) \, \Phi( r ) >$ \ \ with the
angular field:

$$
< \Phi( 0 ) \, \Phi( r ) >
=
\rho_0^2 < e^{ -i { { \bf \phi( 0 ) }
\over \rho_0 } } \ \
e^{ i { { \bf \phi( r ) }
\over \rho_0 } } >
\eqno(4.1)
$$

Divide the angular field operator into creation and
annihilation parts, and commute the former to the left,
the latter to the right, to annihilate the vacuum
states at the respective ends. Then the remaining
commutators give us:

$$
< \Phi( 0 ) \, \Phi( r ) >
=
\rho_0^2 e^{
< { { \bf \phi( 0 ) \, \phi( r ) }
\over \rho_0^2 } > \ -
\ < { { \bf \phi( 0 ) \, \phi( 0 ) }
\over \rho_0^2 } >
}
\eqno(4.2)
$$

which we will have to renormalize due to the
ultraviolet divergence in the second term (the two
terms of the exponent are also infrared divergent
separately, but these divergences cancel each other, so
we can take finite mass, calculate the exponent then
let the mass go to zero). We can renormalize by
dividing the whole expression by \ \ $< \Phi( 0 ) \,
\Phi( r_0 ) >$ , \ where we keep $r_0$ at any fixed
value while letting $r$ go to the event horizon.

$$
{ < \Phi( 0 ) \, \Phi( r ) >
\over
< \Phi( 0 ) \, \Phi( r_0 ) > }
=
e^{
< { { \bf \phi( 0 ) \, \phi( r ) }
\over \rho_0^2 } > \ -
\ < { { \bf \phi( 0 ) \, \phi( r_0 ) }
\over \rho_0^2 } >
}
\eqno(4.3)
$$

As we have seen previously, for \ $r \rightarrow
\alpha$ , \ \ $< \phi( 0 ) \, \phi( r ) >$ \ \ is given
by an infrared divergent integral (it diverges as we
let the mass that regulates the infrared behaviour go
to zero), which divergence also has a \ \ $- \ln( -
\ln( 1 - r^2 / \alpha^2 ) )$ \ \ dependence. From this
follows, that as \ $r \rightarrow \alpha$ , the
exponent, \ \ ${ < { { \phi( 0 ) \, \phi( r ) } \over
\rho_0^2 } > \ - \ < { { \phi( 0 ) \, \phi( r_0 ) }
\over \rho_0^2 } > }$ \ \ will diverge as \ \ $- \ln(
- \ln( 1 - r^2 / \alpha^2 ) )$ \ \ in other words it
will go to \ $-\infty$ . (It might be of interest to
note, that in terms of the conformal radius $\tilde{r}$
this expression for the exponent takes the form (in the
\ $r \rightarrow \alpha$ \ limit): \ \ $- \ln( 2
\tilde{r} / \alpha )$ , \ a single logarithmic
behaviour instead of a double.)

This means, that the \ \ ${ < \Phi( 0 ) \, \Phi( r ) >
\over < \Phi( 0 ) \, \Phi( r_0 ) > }$ \ \ correlation
function goes to zero, as $r$ approaches the event
horizon. This leads us to the conclusion that the
$O(2)$ symmetry of our model is not broken in the
vacuum state of the static system, in other words the
vacuum state of a geodesic observer.

\bigskip

{ \bf 5.Discussion }

\smallskip

In Minkowski spacetime there is not much ambiguity
about the concept of symmetry breaking. The first step,
the choice of the coordinate system for the field
quantization is natural, because the usual rectangular
system has all possible attractive features: all
coordinate lines are geodesics and  translational
invariance is manifest. The only symmetries which need
more complex coordinate-transformations are the
rotational symmetries and Lorentz boosts. However when
we quantize the field, the vacuum state turns out to be
also invariant under these symmetries. It is invariant
even under the Lorentz boosts, although in the
quantization process the equal-time hypersurfaces play
a crucial role at the prescription of the commutation
relations, and these hypersurfaces are certainly not
invariant under Lorentz-boosts. Also, the n-particle
states are mapped into n-particle states, only the
particle 4-momenta transform correspondingly under the
symmetries. Thus all inertial observers have the same
vacuum, and corresponding multi-particle states, which
can be used for studying symmetry breaking, and
possible restoration processes of the symmetric vacuum
state, for example in the case of finite volume.

As we move to a non-flat spacetime, these natural
choices disappear. We have studied field theory in a de
Sitter background, which has constant spacetime
curvature, and a 10-element symmetry (like the
Minkowski symmetry group, which is a contraction of the
de Sitter group). But there is no coordinatization of
the manifold that would consist of geodesics and would
have translational invariance in all of its coordinate
parameters. Furthermore there is no unique vacuum for
all geodesic observers. We have to make our choices for
equal-time surfaces, which will then affect the
quantization through the commutator-prescriptions, that
will lead to the corresponding vacuum states and
Fock-spaces (for discussions of this aspect see for
example [5] and [7]), and finally determine the
existence or absence of symmetry-restoration processes
(which are due to certain infrared divergences,
therefore crucially depend on the properties of the
Fock-space).

Since it is not only a possibility to arrive to
different answers in different coordinate systems, but
indeed the answer was found to be different using
different systems previously [1,4], we think that it is
useful to study the problem through a system that is
selected to correspond to the measurements of a
geodesic observer, leading to a Fock-space where the
vacuum state is the one in which the observer does not
detect particles, etc. (see also [8] and [9]). The
coordinate system that satisfies these requrements is
the ``static'' system of the observer, in which the
spatial sections are geodesic hypersurfaces orthogonal
to the world line of the observer, with the spatial
coordinates translated along the world line of the
observer.

As discussed in [1], the method we have employed
searches for spontaneous symmetry breakdown by
examining \ \ $< \Phi( 0 ) \, \Phi( r ) >$ \ \ as $r$
gets large. If this quantity tends to a positive
constant, it would imply symmetry breaking. On the
other hand if it tends to zero (which is what we have
found), than that is consistent with the restoration
of symmetry. It is still conceivable, however, that
symmetry breaking does take place in this coordinate
system, but manifests itself in a more subtle form.
Studying the structure of the Fock-space might give
more insight into this possibility, as well as into
the physical reasons and the meaning of the different
results that were obtained in refs. [1] and [4] on the
subject of symmetry restoration studied in different
coordinate systems of the same manifold. We intend to
study the problem from this direction in a later paper.

\bigskip

{ \bf Acknowledgements }

\medskip

We are grateful to Hisao Suzuki and Atsushi Higuchi
for helpful discussions.

\hskip.3em
Research supported in part by DOE grant no.
DE-AC02-ERO3075.

\bigskip

{ \bf References }

\bigskip

[1] B. Ratra, Phys. Rev. D {\bf 31} , 1931 (1985).

\smallskip

[2] G. W. Gibbons and S. W. Hawking, Phys. Rev. D
{\bf 15} , 2738 (1976).

\smallskip

[3] B. Allen, Nucl. Phys. B {\bf 226} , 228 (1983).

\smallskip

[4] B. Ratra, Phys. Rev. D {\bf 50} , 5252 (1994).

\smallskip

[5] N. D. Birrell and P. C. W. Davies, {\it Quantum
fields in curved space},

\ \ \ \ Cambridge monographs on mathematical physics,
1985.

\smallskip

[6] A. Higuchi, Class. Quantum Grav. {\bf 4} , 721
(1987);

\ \ \ \ Haru-Tada Sato and Hisao Suzuki, OU-HET 202,
LMU TPW 94-13,

\ \ \ \ hep-th 9410092 \ \ (1994).

\smallskip

[7] E. Mottola, Phys. Rev. D {\bf 31} , 754 (1985).

\smallskip

[8] W. G. Unruh, Phys. Rev. D {\bf 14} , 870 (1976).

\smallskip

[9] A. Z. Capri and S. M. Roy, Int. Journ. of Mod.
Phys. A  {\bf 9} , 1239 (1994).

\smallskip

[10] M. Abramowitz and I. A. Stegun, {\it Handbook of
mathematical functions},

\ \ \ \ \ Dover books on advanced mathematics, 1972.

\end